\documentclass[12pt,ngerman,american]{iopart}

\usepackage[utf8]{inputenc}         
\usepackage[main=american]{babel} 
\usepackage[T1]{fontenc}            
\usepackage{cite}

\usepackage[dvipsnames]{xcolor}         
\usepackage[%
  unicode=false,%
  pdftitle={A primer to numerical simulations: The perihelion motion of Mercury},%
  pdfauthor={C. K\"orber, I. Hammer, J.-L. Wynen, J. Heuer, C. M\"uller and C. Hanhart},%
  pdfkeywords={Numerical Simulations, Teaching, Perihelion Motion of Mercury, General Relativity},%
  pdfsubject={A primer with the goal to teach general concepts of numerical simulations to high school students.  The specific example is the perihelion motion of Mercury due to General Relativity.},   
  colorlinks=true,        
  linkcolor=OliveGreen,          
  citecolor=magenta,        
  filecolor=YellowGreen,      
  urlcolor=blue           
]{hyperref}
\usepackage[space=true]{accsupp}
\newcommand{\copyablespace}{\BeginAccSupp{method=hex,unicode,ActualText=00A0}\ \EndAccSupp{}}

\usepackage{textcomp,gensymb}       

\usepackage[autostyle=true]{csquotes}            
\MakeOuterQuote{"}                               

\usepackage{amssymb}                  
\usepackage{graphicx}                 
\usepackage{listings}                 

\usepackage{subcaption} 

\definecolor{fzjblue}{HTML}{005B81}
\definecolor{light-gray}{gray}{0.97}
\definecolor{semilight-gray}{gray}{0.67}
\definecolor{mygreen}{rgb}{0,0.6,0}

\usepackage{tikz}
\usetikzlibrary{shapes.geometric, arrows}
\tikzstyle{startstop} = [
	rectangle,
	rounded corners,
	minimum width=3cm,
	minimum height=1cm,
	text width=4.5cm,
	text centered,
	draw=black,
	fill=green!30
]

\tikzstyle{process} = [
	rectangle,
	minimum width=3cm,
	minimum height=1cm,
	text width=4.5cm,
	text centered,
	draw=black,
	fill=blue!25
]
\tikzstyle{decision} = [
	diamond,
	minimum width=4.5cm,
	minimum height=1cm,
	text width=1.7cm,
	text centered,
	draw=black,
	fill=red!30
]
\tikzstyle{arrow} = [thick,->,>=stealth]

\newcommand{\python}[0]{\texttt{Python}}
\newcommand{\vpython}[0]{\texttt{VPython}}
\newcommand{\abs}[1]{\left\vert #1 \right\vert}
\newcommand{\code}[1]{{\scriptsize\colorbox{light-gray}{\texttt{#1}}}}

\begin{document}

\title[A primer to numerical simulations]{A primer to numerical simulations: The perihelion motion of Mercury}

\author{
	C.~K\"orber$^{1}$,
	I.~Hammer$^{1}$,
	J.-L.~Wynen$^{1}$,
	J.~Heuer$^{2}$\footnote{Present Address: Institut f\"ur Neurowissenschaften und Medizin (INM-4), Forschungszentrum J\"ulich, D-52425 J\"ulich, Germany},
	C.~M\"uller$^{3}$ and
	C.~Hanhart$^{1}$
}
\address{
	$^1$ \textit{Institut f\"ur Kernphysik (IKP-3) and Institute for Advanced Simulations (IAS-4), Forschungszentrum J\"ulich, D-52425 J\"ulich, Germany}\\
	$^2$ \textit{Hochschule Hamm-Lippstadt, Marker Allee 76-78, 59063 Hamm, Germany}\\
	$^3$ \textit{Sch\"ulerlabor JuLab, Forschungszentrum J\"ulich, D-52425 J\"ulich, Germany}
}
\ead{c.koerber@fz-juelich.de, c.hanhart@fz-juelich.de}
\vspace{10pt}

\begin{abstract}
Numerical simulations are playing an increasingly important role in modern science.
In this work it is suggested to use a numerical study of the famous perihelion motion of the planet Mercury (one of the prime observables supporting Einsteins General Relativity) as a test case to teach numerical simulations to high school students.
The paper includes details about the development of the code as well as a discussion of the visualization of the results.
In addition a method is discussed that allows one to estimate the size of the effect as well as the uncertainty of 
the approach a priori. At the same time this enables the students to double check the results found numerically.
The course is structured into a basic block and two further refinements which aim at more advanced students.
\end{abstract}

%
%
%
%
%

\section{Introduction}\label{sec:intro}

Numerical simulations play a key role in modern physics because they allow one to tackle some theoretical problems not accessible otherwise. This might be the case because there are too many particles participating in the system (as in simulations for weather predictions) or the interactions are too complicated to allow for a systematic, perturbative approach (as in theoretical descriptions of nuclear particles at the fundamental level).
This paper introduces a project that allows one to demonstrate the power of numerical simulations.
On the example of the perihelion motion of the planet Mercury the students are supposed to learn about
\begin{itemize}
\item the importance of differential equations in theoretical physics;
\item the numerical implementation of Newtonian dynamics;
\item systematic tests and optimization of computer codes;
\item effective tools to estimate the result a priori as an important cross check;
\item the visualization of numerical results using \vpython{}.
\end{itemize}
We are convinced that in order to excite students for numerical simulations it is compulsory to demonstrate their power on an example that catches their interest.
This purpose is served perfectly by the case chosen here, since Einsteins equations of General Relativity are fascinating to a very broad public.
Their detailed study needs a deep understanding of Differential Geometry. For the course presented here, however, very little math is necessary, such that high school students vaguely familiar to vector calculus and derivatives will benefit from it.
This was already demonstrated in the "Sch\"ulerakademie Teilchenphysik", where this course was tested successfully on two groups consisting in total of 24 German high school students from 10$^{\rm th}$ to 13$^{\rm th}$ grade in 2015 and 2017.

The course as well as this paper is structured as follows: after an introduction to Newtonian dynamics and the concept of differential equations, their
discretization is discussed based on Newton's law of gravitation and possible extensions thereof.
Afterwards the visualization of the resulting trajectories using \vpython{} is introduced and applied to the problem at hand. In addition, tools are developed to extract the relevant quantity from the result of the simulation.
Finally, the principle of dimensional analysis is presented as a tool to estimate the size of the effect studied as well as its expected accuracy.
Furthermore the method allows one to cross check the results of the simulation.
As is pointed out in the chapters below the course can be finished at some canonical points --- in particular we think
that for most students a detailed study of the uncertainties of the numerical simulations might be too technical.
However, we include this discussion as well both for completeness and as an additional challenge to the more advanced students.

\section{Trajectories, velocities, accelerations and Newton's second law}\label{sec:tva}

A physical system is said to be understood, if the assumed forces acting on it lead to the observed trajectories.
In other words, we need to show that we can calculate the location in space of the object of interest at any future point in time, once the initial conditions are fixed properly.
If we can neglect the finite size of this object (and in particular its orientation in space), the location is parametrized by a single three vector $\vec r(t)$.
Additionally, we need to be able to calculate the objects velocity $\vec v(t)$ and acceleration $\vec a(t)$ in order to describe and control its dynamics; this will become clear below. The velocity describes a change of location over time which, for some infinitesimally small $\Delta t$, can be written as
\begin{equation}\label{eq:x_update}
\vec r(t+\Delta t) = \vec r(t) + \vec v(t) \Delta t + \ldots \ . \label{eq:vdef}
\end{equation}
Similarly, the acceleration describes a change in velocity, i.e.
\begin{eqnarray}\label{eq:v_update}
\vec v(t+\Delta t) = \vec v(t) + \vec a(t) \Delta t + \ldots\ . \label{eq:adef}
\end{eqnarray}
The dots in these expressions indicate additional terms that may be expressed in terms of higher powers in $\Delta t$. While they are needed in general, for sufficiently small $\Delta t$ they can be safely neglected.
Thus we may define the time derivative via
\begin{equation}
\vec v(t) = \lim_{\Delta t\to 0} \frac{\Delta \vec r(t)}{\Delta t} =: \frac{d\vec r(t)}{d t}  = \dot{\vec  r}(t) \ ,
\end{equation}
where $\Delta \vec r(t)=\vec r(t+\Delta t)-\vec r(t)$ and we introduced a common short hand notation for time derivatives in the last expression.
Analogously we get
\begin{eqnarray}
\vec a(t) &=& \lim_{\Delta t\to 0} \frac{\Delta \vec v(t)}{\Delta t} =: \frac{d\vec v(t)}{dt}  = \dot{\vec v}(t) \ , \\
 &=& \frac{d^2\vec r(t)}{dt^2}  = \ddot{\vec r}(t) \ ,
\end{eqnarray}
where we introduced the second derivative in the second line.

It was Newton who observed that, if a body is at rest, it will remain at rest, and if it is in motion it will remain in motion at a constant velocity in a straight line, 
unless it is acted upon by some force --- this is known as Newton's first law.
Formulated differently: a force $\vec F$ expresses itself by changing the motion of some object.
This is quantified in Newton's second law
\begin{equation}
\vec F(\vec r, t) = \frac{d}{dt}(m \vec v) \ .
\end{equation}
If the mass does not change~\footnote{%
	A well known example where $m$ does change with time is a rocket, whose mass decreases as fuel is burned.
} with time, this reduces to the well known
\begin{equation}
\vec F(\vec r) = m \vec a(t) = m\dot{\vec v}(t) = m\ddot{\vec{r}}(t) \ . \label{eq:newton2}
\end{equation}
Note that in general the force could depend also on the time or the velocity.
Here, we restrict ourselves to the case relevant for our example, where the force depends on the location only.
Therefore, as soon as the force $\vec F(\vec r)$ is known for all $\vec r$ one can in principle calculate the trajectory by solving Eq.~(\ref{eq:newton2}) for $\vec r(t)$.
Sometimes this requires some advanced knowledge of mathematics and sometimes no closed form solution exists.
However, alternatively one can calculate the whole trajectory of some test body that experiences this force by a successive application of the rules given in Eqs.~(\ref{eq:vdef}) and (\ref{eq:adef}):
\begin{enumerate}
\item For a given time $t$, where $\vec r(t)$ and $\vec v(t)$ are known, use the force to calculate $\vec a(t) = \vec F(\vec r(t)) / m$, see Eq.~(\ref{eq:newton2}).
\item Use Eq.~(\ref{eq:vdef}) to calculate $\vec r(t+\Delta t)$.
\item Use Eq.~(\ref{eq:adef}) to calculate $\vec v(t+\Delta t)$.
\item Go back to (i) with $t\to t+\Delta t$.
\end{enumerate}
Clearly to initiate the procedure at some time $t_0$ both $\vec r(t_0)$ as well as $\vec v(t_0)$ must be known --- the  trajectories depend on these initial conditions~\footnote{%
	In general, a differential equation of $n^{\rm th}$ degree (where the highest derivative is order $n$) needs $n$ initial conditions specified.
	For $n=2$ those are often chosen as location and velocity at some starting time, but one may as well pick two locations at different times.%
}.

This procedure can only work if $\Delta t$ is sufficiently small.
One way to estimate whether $\Delta t$ is small enough is to verify whether the relation
\begin{equation}
|\vec v(t)| \gg \frac12|\vec a(t)|\Delta t = \frac{1}{2m} |\vec F(\vec r(t)) |\Delta t\
\label{eq:check}
\end{equation}
holds. This follows from Eq.~(\ref{eq:vdef}) where the first term that we neglected reads $(1/2)a(t){(\Delta t)}^2$.
Eq.~({\ref{eq:check}}) also shows that small (large) time steps are necessary (sufficient), if the force is strong (weak), since the time steps need to be small enough that all changes induced by the force get resolved.
Clearly, a relation as Eq.~({\ref{eq:check}}) can only provide guidance and can not replace a careful numerical check of the solutions: A valid result has to be insensitive to the concrete value chosen for $\Delta t$ --- in particular replacing $\Delta t$ by $\Delta t/2$ should not change the result significantly.
This issue will be discussed in more detail below.

\section{Example: General Relativity and the perihelion motion of Mercury}\label{sec:gr}
For this concrete example the starting point for the force is Newton's law of gravitation
\begin{equation}
\vec F_N(\vec r) = - \frac{G_N m M_\odot}{r^2} \frac{\vec r}{r}\ ,
\end{equation}
where $G_N=6.67\times 10^{-11}$ m$^3$kg$^{-1}$s$^{-2}$ is the Newtonian constant of gravitation, $m$ is the mass of Mercury and $M_\odot=1.99\times 10^{30}$ kg is the mass of the Sun.
In addition $r=|\vec r(t)|$ denotes the distance between Sun and Mercury, when we assume that the Sun is infinitely heavier than Mercury and located at the center of the coordinate system.
Although this is not exact, it is a good approximation because $m/M_\odot\sim 10^{-8}$.
For later convenience we introduce the Schwarzschild radius of the Sun
\begin{equation}
r_S=\frac{2G_N  M_\odot}{c^2} = 2.95 \ \mbox{km} \ , \label{rsdef}
\end{equation}
where $c=3.00\times 10^8$ m/s denotes the speed of light.
Note that $r_S$ is the characteristic length scale of the gravitational field of the Sun for --- up to the prefactor --- one can not form another quantity with dimensions of a length from $G_N$, $M_\odot$ and $c$.
The Schwarzschild radius is also an important quantity to characterize black holes; this is however not relevant for the discussion at hand.
With this, Newton's second law reads
\begin{equation}
\ddot{\vec r}      = - \frac{c^2}{2}\left(\frac{r_S}{r^2}\right)\frac{\vec r}{r} \, . \label{eq:newton}
\end{equation}
An attractive force that vanishes at large distances leads in general, depending on the initial conditions, either to bounded or to open orbits.
In the latter case the planet simply disappears from the Sun, since a given force can capture only bodies with small enough momenta.
The students may study those scenarios within their simulations by varying the start velocity while keeping the start location fixed.

The bounded orbits that emerge from a potential that scales as $1/r$ (which corresponds to a force that scales as $1/r^2$) are elliptic and fixed in space.
In particular the point of closest approach of the planet to the Sun, the perihelion, does not move.
However, when a potential that vanishes for $r\to \infty$ deviates from $1/r$, the perihelion does move.
This makes the behavior of the perihelion a very sensitive probe of the gravitational potential.

The observed perihelion motion of Mercury is nowadays determined as
\[(574.10\pm 0.65)'' \ \mbox{per 100 earth years \cite{RevModPhys.19.361}}\]
where the symbol $''$ denotes "arc seconds": 1$''={(1/3600)}^o$. The bulk of this number
can be understood by the presence of other planets within the Newtonian theory, since
their gravitational force also acts on Mercury. However, a
residual motion of
\begin{equation}
\delta \Theta_M = (42.56\pm 0.94)'' \ \mbox{per 100 earth years \cite{RevModPhys.19.361}} \label{delT}
\end{equation}
remained unexplained, until Einstein quantified the predictions of General Relativity to this particular observable \cite{Einstein}
\begin{equation}
\delta \Theta_{GR} = (43.03\pm 0.03)'' \ \mbox{per 100 earth years \cite{RevModPhys.19.361}} \label{delTGR} \, .
\end{equation}

To allow for a movement of the perihelion we need to modify the force that led to Eq.~(\ref{eq:newton}).
This modification should be such that it depends on $r$ and it vanishes for $r\to \infty$.
Based on the discussion presented above we may multiply the right hand side of~(\ref{eq:newton}) with a factor
$(1+\alpha \, \frac{r_S}{r} )$, where $\alpha$ denotes some dimensionless parameter.
This seems like a natural way to parametrize the additional potential because it adds the smallest possible deviation from a $1/r$ potential expressed relative to the characteristic length $r_S$.
However, besides $r_S$, which characterizes the gravitational potential of the Sun, there is also a characteristic parameter for the dynamics of Mercury:
\begin{equation}
r_L^2 := \frac{\vec L\,^2}{m^2c^2}= \frac{{(\vec r\times \dot{\vec r} \, )}^2}{c^2} \ , \label{a2def}
\end{equation}
where $\vec L$ denotes the angular momentum, which is a constant of motion for central potentials.
It is easily verified that $r_L^2$ carries dimensions of length squared.
We therefore have to add one more term to the potential; in analogy to the one discussed above, we use $\beta \, \frac{r_L^2}{r^2}$.
Here one might wonder why this additional term is chosen proportional to $(r_L/r)^2$ and not to $(r_L/r)$. The reason for this choice
is indeed not obvious: In general the correction terms added must be scalar quantities --- accordingly vectors can enter only
as scalar products. Moreover, one is not allowed to use square-roots of those scalar products for these impose
wrong mathematical properties to the equations of motion (for the case at hand, e.g., the derivative with
respect to the velocity would not be defined at $\dot{\vec r}=0$) --- the only vector that is allowed to appear as its length in linear order
is the radius, since this length is related to geometrical properties of the system.  
Combining everything, we use the following ansatz for the modified equation of motion:
\begin{equation}
\ddot{\vec r} = - \frac{c^2}{2}\frac{r_S}{r^2}\left(1+\alpha\frac{r_S}{r}+\beta\frac{r_L^2}{r^2}\right) \, \frac{\vec{r}}{r} \ .
\label{eq:newton_art}
\end{equation}
For the system at hand, $r_L^2$ may be estimated from the parameters of Mercury at its perihelion: The corresponding velocity is
 $\dot{ r}(t=0)=\abs{\vec v_M(0)} = 59.0$~km/s (here we already indicate that we will start the simulation at the perihelion) and the closest distance between Sun and Mercury is
$ r_{MS}=\abs{\vec r_{MS}(0)} = 46.0 \cdot 10^6$~km~\cite{MercuryFactSheet}. Note that at the perihelion $\vec r$ and $\dot{\vec r}$ are perpendicular to each other (see Figure \ref{fig:sun_merc}).
We thus find that $(r_L^2/ r_{MS}^2) \sim 4\cdot10^{-8}$. Furthermore $(r_S/r_{MS})\sim6\cdot 10^{-8}$.
Therefore one may expect from both correction terms an effect of similar size --- for a more detailed discussion about the underlying logic we refer to Sec.~\ref{sec:analysis}.
Furthermore, any term of higher order in either $r_L^2$ or $r_S$
should be suppressed by additional seven orders of magnitude and thus can be neglected safely.
The parameters $\alpha$ and $\beta$ can be extracted from a fit to data.
However, since we have only a single number to study, the perihelion shift (see Eq.~(\ref{delT})),
we can only fix either $\alpha$ or $\beta$ (or a linear combination thereof).
The parameters can also be calculated from the underlying theory one uses. The actual values depend on the specific theory;
General Relativity gives~\cite{Einstein}
\begin{equation}
	\alpha = 0 \, , \quad \beta = 3 \, . \label{eq:art-prediction}
\end{equation}
Thus we may also use the simulation explained below to calculate the perihelion motion of Mercury from the input
values given in Eq.~(\ref{eq:art-prediction}).
Modifications to the Newtonian equation of gravity introduced to account for the perihelion motion of Mercury
are discussed in great detail in a very pedagogical way also in Ref.~\cite{Wells:2011st}.


\section{Numerical Implementation}\label{sec:Numerical Implementation}

\subsection{Describing the Motion with Python}

In this section we present the numerical implementation as well as the visualization of planetary trajectories and in
particular the perihelion motion of Mercury.
We choose \python{}~\cite{Python} as programming language because \python{} is easy to learn, intuitive to understand and an open source language.
No prior knowledge of \python{} or any other programming language is required, as we explain all necessary steps to create the simulation.
In the following, we present code examples which tested for \python{} version \texttt{3.6.4} and \vpython{} version \texttt{7.3.2}.
We provide online instructions for setting up \python{} and \vpython{} on different operating systems \cite{scripts}.
In \ref{sec:code}, we also provide a working example as well as possible extensions and template files online.

To start the simulation one needs the "initial" distance and velocity [Eq.~(\ref{eq:x_update}) and Eq.~(\ref{eq:v_update})].
As described above, here one can simplify the problem by placing one object (the "infinitely" heavy Sun) in the center of the coordinate system and keep it fixed.

Since the trajectories do not depend on where on the orbit of Mercury the simulation is started, we
use the values at the perihelion with $\abs{\vec r_{MS}(0)} = 46.0 \cdot 10^6\,\mathrm{km}$ and $\abs{\vec v_M(0)} = 59.0\,\mathrm{km/s}$\cite{MercuryFactSheet} as initial (for $t=0$) parameters.
The computer does not understand physical units, hence one has to express each variable in an appropriate unit.
Both for the numerical treatment and the intuitive understanding, it is useful to select parameters in a "natural range", e.g.,
by expressing distances in $R_0 = 10^{10}\,\mathrm{m}$ and time intervals in days, $T_0 = 1\,\mathrm{d}$, where d refers to earth days. One Mercury year is given by $T_M=88.0\,T_0$.
With this choice, the initial distance of Mercury to the Sun, the size of the initial velocity of Mercury and the acceleration prefactor (see Eq.~(\ref{eq:newton_art})) become
\begin{eqnarray}
r_{MS}(0)   &=& 4.60 R_0 \, , \quad
v_{M}(0)    = 0.510 \frac{R_0}{T_0} \, ,  \\
a_{MS}(r_{MS}) &=& \frac{c^2}{2}\frac{r_S}{r_{MS}^2} = 0.990 \frac{R_0}{T_0^2} \frac{1}{{\left(r_{MS}/R_0\right)}^2}\, .
\end{eqnarray}

\begin{figure}[htb]
	\centering
	\includegraphics[width=.5\textwidth]{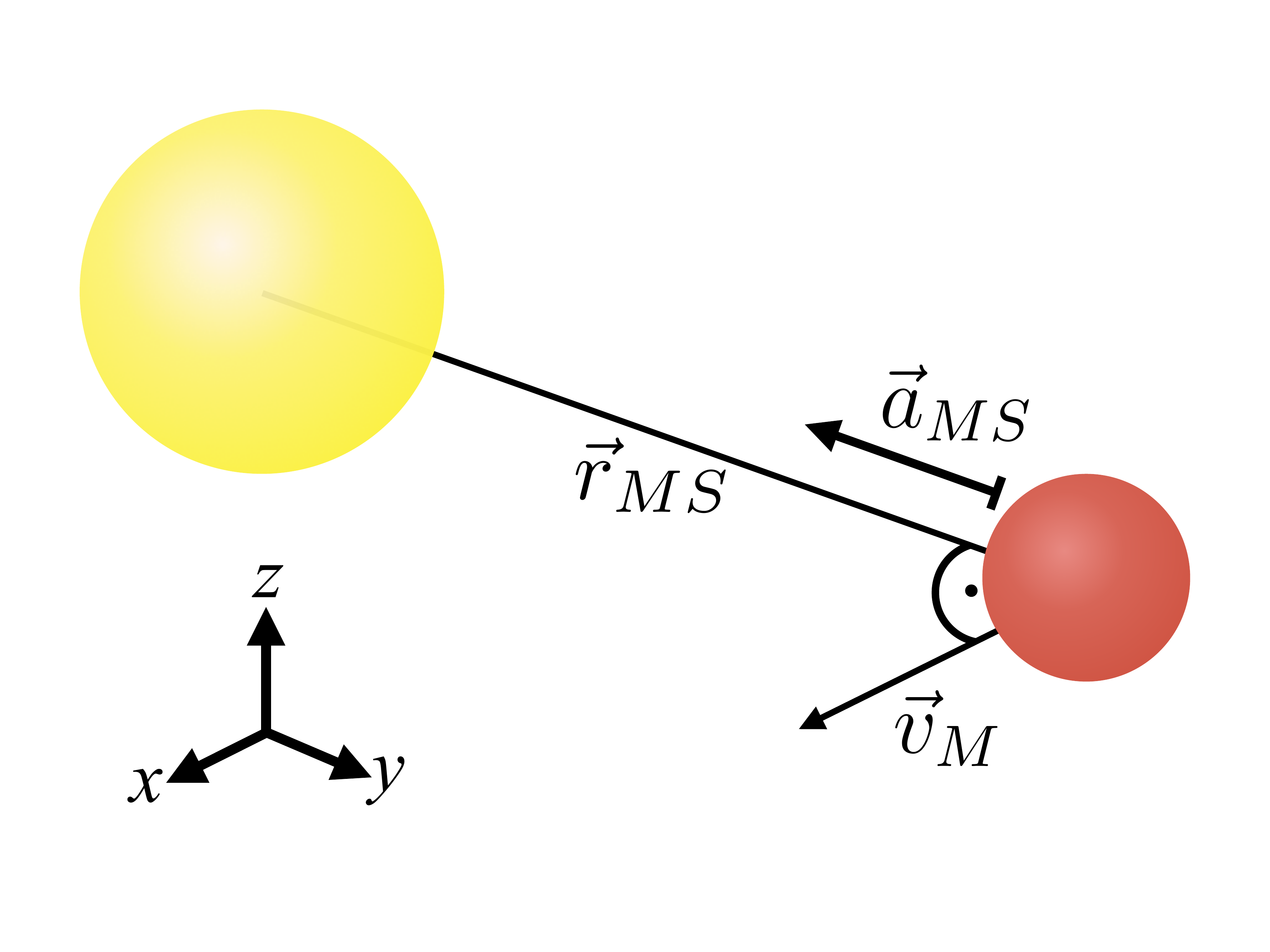}
	\caption{\label{fig:sun_merc}%
		Sun-Mercury system with relevant vectors.
		Mercury is at its perihelion, therefore its velocity is perpendicular to its direct connection vector with the sun.%
	}
\end{figure}

In \python{}, this reads
\begin{lstlisting}
# Definition of parameters
rM0 = 4.60    # Initial radius of Mercury orbit, in units of R0
vM0 = 5.10e-1 # Initial orbital speed of Mercury, in units of R0/T0
c_a = 9.90e-1 # Base acceleration of Mercury, in units of R0**3/T0**2
TM  = 8.80e+1 # Orbit period of Mercury
rS  = 2.95e-7 # Schwarzschild radius of Sun,in units of R0
rL2 = 8.19e-7 # Specific angular momentum, in units of R0**2
\end{lstlisting}
where $\texttt{c\_a} =a_{MS} r_{MS}^2$, the acceleration without the radius factor.
The last two quantities refer to the Schwarzschild radius of the Sun and the parameter $r_L^2$ defined in Eqs. (\ref{rsdef}) and (\ref{a2def}), respectively.

So far we only fixed the length of the vectors. Next, we set up the initial directions, which will describe the motion in space.
We will build on the existing \python{} module \vpython{}, which provides an implementation for treating vectors as well as their visualization.
The first object of interest is a \texttt{vector}, which takes three-dimensional coordinates as its input.
With our choice of initial conditions (we picked the initial vectors in the perihelion), the velocity of Mercury is perpendicular to the vector which connects Mercury and Sun (see figure~\ref{fig:sun_merc}):
\begin{lstlisting}
# Import the class vector from vpython
from vpython import vector
# Initialize distance and velocity vectors
vec_rM0 = vector(0, rM0, 0)
vec_vM0 = vector(vM0, 0, 0)
\end{lstlisting}\footnote{Note that the code is designed for \vpython{} versions \texttt{7} and later. For \vpython{} versions prior \texttt{7}, e.g., the statement from \code{vpython import vector}  must be replaced by \code{from visual import vector}.}
We use the estimate of Eq.~(\ref{eq:check}) as a guidance to fix the time step $\Delta t$ as
\begin{lstlisting}
# Definition of the time step
dt = 2 * vM0 / c_a / 20
\end{lstlisting}
Here the factor $1/20$ makes sure that $\Delta t$ is indeed consistent with Eq.~(\ref{eq:check}).
Now we are in the position to calculate location and velocity of the planet at $t_0+\Delta t$ using the following expressions\footnote{%
	Note, that there are different notions for numerically integrating differential equations with different accuracies.
	Thus, the ordering and exact expressions for updating $a_{MS}$, $v_M$ and $r_{MS}$ are not unique. However, all correct prescriptions lead to the same result for sufficiently small $\Delta t$.
	For this particular problem, the definitions we show are in good balance between being numerically stable and inexpensive to compute.
}%
\begin{lstlisting}
# Compute the strength of the acceleration
temp = 1 + alpha * rS / vec_rM_old.mag + beta * rL2 / vec_rM_old.mag**2
aMS  = c_a * temp / vec_rM_old.mag**2
# Multiply by the direction
vec_aMS = - aMS * ( vec_rM_old / vec_rM_old.mag )
# Update velocity vector
vec_vM_new = vec_vM_old + vec_aMS * dt
# Update position vector
vec_rM_new = vec_rM_old + vec_vM_new * dt
\end{lstlisting}
Note basic vector operations are already implemented in the predefined \texttt{vector} class.
The difference and sum of two vectors, or the scalar vector multiplication return vectors themselves.
Also the magnitude of a vector --- \code{vector.mag} --- is an attribute of the vector and can be easily extracted.

It is handy to use \texttt{Python}s functions to structure the program and hold repeating code ("DRY" --- Don't Repeat Yourself):
\begin{lstlisting}
# Define the coordinate and velocity update function
def evolve_mercury(vec_rM_old, vec_vM_old, alpha, beta):
    # Compute the strength of the acceleration
    temp = 1 + alpha * rS / vec_rM_old.mag + beta * rL2 / vec_rM_old.mag**2
    aMS  = c_a * temp / vec_rM_old.mag**2
    # Multiply by the direction
    vec_aMS = - aMS * ( vec_rM_old / vec_rM_old.mag )
    # Update velocity vector
    vec_vM_new = vec_vM_old + vec_aMS * dt
    # Update position vector
    vec_rM_new = vec_rM_old + vec_vM_new * dt
    return vec_rM_new, vec_vM_new

# Call the function
vec_rM_new, vec_vM_new = evolve_Mercury(vec_rM_old, vec_vM_old, 0.0, 0.0)
\end{lstlisting}
\python{}s syntax enforces a clean programming style: it is necessary that the body of the function is indented (by an arbitrary but consistent amount of \texttt{space}s\footnote{While \python{} allows using \texttt{tab}s as well, this can cause programming errors, because their displayed width depends on the text editor.}) relative to the definition statement of the function.
Furthermore, \python{} is an Interpreter language.
Each line of the code is executed when the Interpreter passes it.
For this reason, if we define the parameters before the function, they can be used inside of the function.
Variables defined within the function (e.g. \code{aMS}, \code{vec\_aMS}, $\dots$) are local and do not exist beyond the scope of the function; they can not be used after the \texttt{return} statement.

Finally, we can describe the evolution by a \texttt{while}-loop
\begin{lstlisting}
t     = 0.0
alpha = 0.0
beta  = 0.0
# Set position and velocity to their starting points
vec_rM = vec_rM0
vec_vM = vec_vM0
# Execute the loop as long as t < 2*TM
while t < 2*TM:
    # Update position and velocity
    vec_rM, vec_vM = evolve_mercury(vec_rM, vec_vM, alpha, beta)
    # Advance time by one step
    t = t + dt
\end{lstlisting}
where for the start we set the parameters $\alpha=0 = \beta$ in order to first study the properties of the pure
$1/r^2$ force.
Note the required indent of the loop structure similar to the indent of a function.
In each iteration of the \texttt{while}-loop, the previous distance and velocity are used to compute the new values,
directly overwriting the previous values.
The total runtime \texttt{2*TM} is the amount of "virtual" days the simulations should run.
To describe at least one full orbital period, this time needs to be larger than $T_M$.
With the previous choice for \texttt{dt}, this corresponds to roughly $N_T \approx 2 \cdot 10^3$ evolution steps.
Note that the exact time it takes for Mercury to complete a full revolution depends on the initial coordinates and velocities, the accuracy of the computation
(controlled by the value of $\Delta t$) as well as the computing power employed.
One should encourage the students to analyze this in the beginning.

\subsection{Visualizing the Motion with VPython}
To start the visualization one has to \texttt{import} further objects from the \texttt{VPython} module.
We change the import statement from before to
\begin{lstlisting}
from vpython import vector, sphere, color, curve, rate
\end{lstlisting}
The class sphere will represent Mercury and the Sun in the simulation
\begin{lstlisting}
# Define the initial coordinates; M = Mercury, S = Sun
M = sphere(pos=vec_rM0,       radius=0.5,  color=color.red   )
S = sphere(pos=vector(0,0,0), radius=1.5,  color=color.yellow)
# And the initial velocities
M.velocity = vec_vM0
S.velocity = vector(0,0,0)
# Add a visible trajectory to Mercury
M.trajectory = curve(color=color.white)
\end{lstlisting}
We place the Sun in the origin of our coordinate system and choose non-realistic radii for visualization purposes.
Last but not least, one should use in the code the vectors directly related to the visualization. Thus
the \texttt{while}-loop becomes
\begin{lstlisting}
t     = 0.0
alpha = 0.0
beta  = 0.0
# Execute the loop as long as t < 2*TM
while t < 2*TM:
    # Set the frame rate (you can choose a higher rate to accelerate the program)
    rate(100)
    # Update the drawn trajectory with the current position
    M.trajectory.append(pos=M.pos)
    # Update velocity and position
    M.pos, M.velocity = evolve_mercury(M.pos , M.velocity , alpha, beta)
    # Advance time by one step
    t = t + dt
\end{lstlisting}

\begin{figure}[htb]
	\centering
	\begin{subfigure}[c]{0.22\textwidth}
		\includegraphics[width=\textwidth]{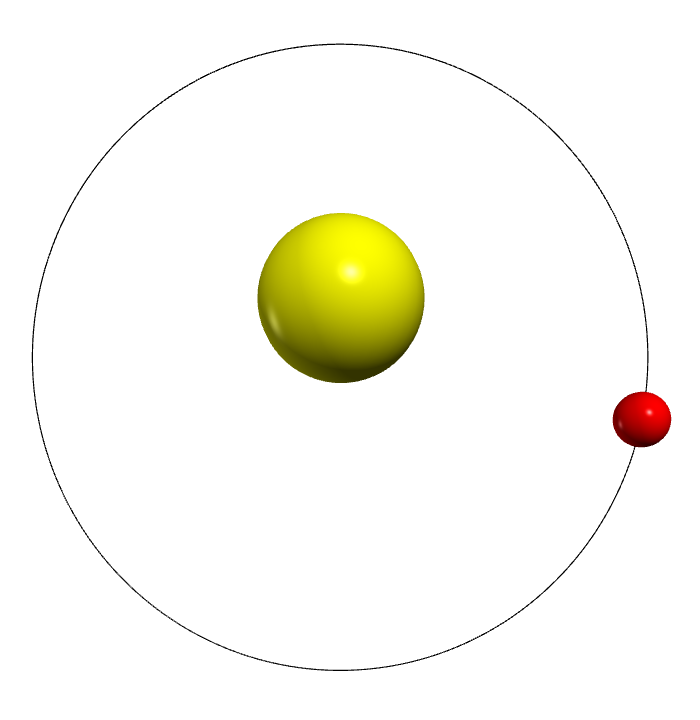}
		\caption{\label{fig:MercuryOrbit-a0-small}}
	\end{subfigure}
	~
	\begin{subfigure}[c]{0.22\textwidth}
		\includegraphics[width=\textwidth]{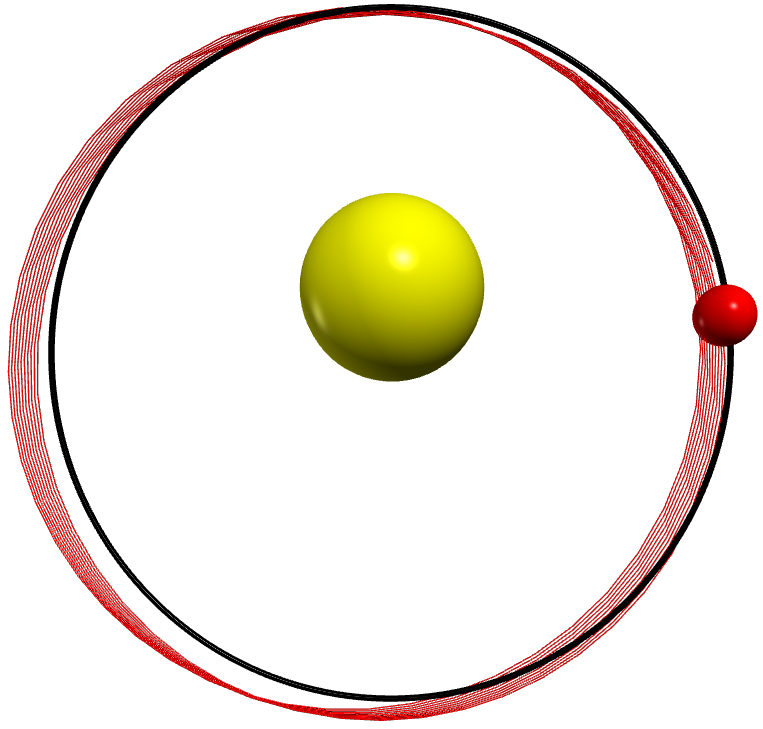}
		\caption{\label{fig:MercuryOrbit-a0-small-dt-large}}
	\end{subfigure}
	~
	\begin{subfigure}[c]{0.22\textwidth}
		\includegraphics[width=\textwidth]{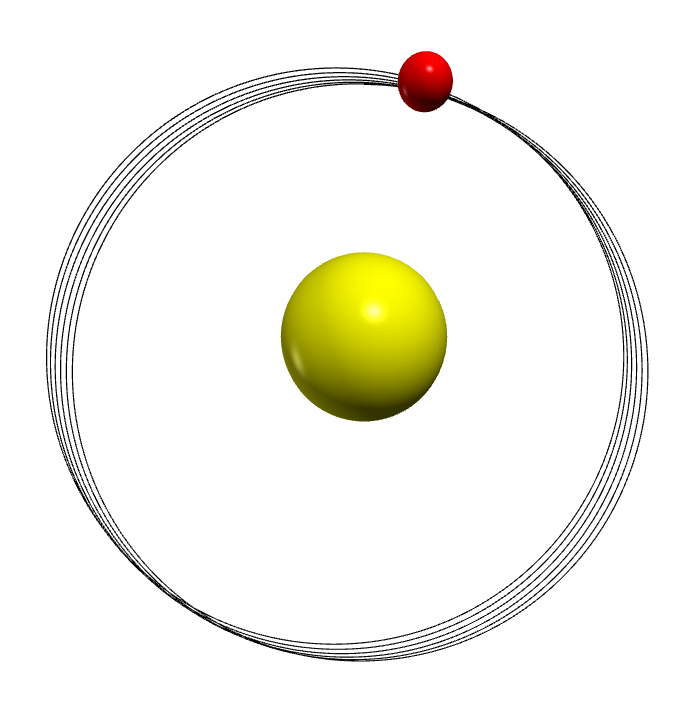}
		\caption{\label{fig:MercuryOrbit-a6-small}}
	\end{subfigure}
	~
	\begin{subfigure}[c]{0.22\textwidth}
		\includegraphics[width=\textwidth]{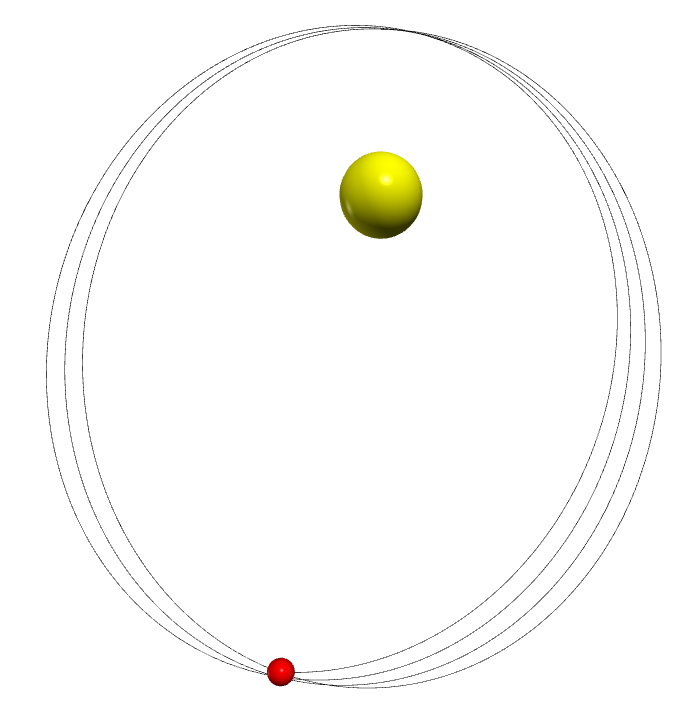}
		\caption{\label{fig:MercuryOrbit-a6-big}}
	\end{subfigure}
	\captionsetup{singlelinecheck=off}
	\caption[]{\label{fig:MercuryOrbit}
		Different Mercury orbits for $\beta=0$ and
		\begin{itemize}
		\item[(a)] $\alpha=0$ and $\Delta t = \Delta t_0  / 20$;
		\item[(b)] $\alpha=0$ and $\Delta t = \Delta t_0  \times 2$, where the black line represents the orbit of (a) and the red line is the orbit for the larger time steps;
		\item[(c)] $\alpha=10^{6}$ and $\Delta t = \Delta t_0 / 20$;
		\item[(d)] $\alpha=10^{6}$ and $\Delta t = \Delta t_0 / 20$ but $r_{MS}(0) = 6 R_0$.
		\end{itemize}
		Here, the time steps are defined by $\Delta t_0 \equiv 2 v_{M}(0) /a_M(0)$ and the images are screenshots of the simulation (with modified colors).
		Note that for different starting values of $r_{MS}(0)$, also the numbers of days for a "Mercury year" changes.
	}
\end{figure}

If the starting values are chosen as advised in the previous section, the students should end up with a trajectory as depicted in Figure~\ref{fig:MercuryOrbit-a0-small}.
At this point one might ask students to vary $\Delta t$ and observe the effect of this on the orbit [see also Figure~\ref{fig:MercuryOrbit-a0-small-dt-large}].
To get the perihelion motion, the additional force term described in Eq.~(\ref{eq:newton_art}) has to be "turned on", e.g., by choosing either $\alpha\neq0$ or $\beta\neq0$
[Figure~\ref{fig:MercuryOrbit-a6-small}].
This is a good opportunity to let the students play with the size of $\alpha$ or $\beta$ and get a feeling for its impact on the trajectories. In particular
they should discover that values of $\alpha$ or $\beta$ larger than $10^5$ are necessary to get a visible effect.
For an even more enhanced perihelion motion it is advisable to not use the correct Mercury values but a more excentric trajectory by choosing, e.g., $r_{MS}(0)=6R_0$ [Figure~\ref{fig:MercuryOrbit-a6-big}].

This finalizes the basic course ---
for the not so ambitious students this might be a good point to stop. Already up to here the students should have
learned a lot about numerical physics by successfully simulating the perihelion motion. The chapters to come are increasingly technical
and thus address more advanced students.

\subsection{Extracting the Perihelion Motion}

Both $\alpha$ and $\beta$ have similar effects on the perihelion motion.
We recommend to set one of them to zero and to vary the other, when analyzing the perihelion motion. 
In order to calculate the size of the perihelion motion for a modified gravitational force, the students need to
\begin{itemize}
\item extract multiple positions of the perihelion $\vec r_{MS} ( t_\mathrm{ph}^{(n)} )$ for a fixed value of $\alpha$ and $\beta$. The  simulation time needs to cover several revolutions.
\item calculate the angle between the perihelions for each pair of successive turns and compute the average angle $\delta \Theta (\alpha)$ over all individually computed angles.
\item repeat the steps given above for different values of $\alpha$ (or $\beta$). The students should find a linear dependence of $\delta \Theta$ on $\alpha$ or $\beta$.
\item interpolate $\delta \Theta (\alpha, \beta)$ to small values of $\alpha$ (or $\beta$) --- including  $\delta \Theta(0,0)=0$.
\end{itemize}

There are several ways to find the position of the perihelion.
The easiest one is to look for the point of minimal distance to the sun --- the definition of the perihelion.
This can be done within the $while$-loop.
To identify a minimal distance to the sun, one has to know the last two positions \code{vec\_rM\_past} and \code{vec\_rM\_before\_last} in addition to the current position.
If the length of the \texttt{last} vector is smaller than that of the \texttt{before\_last} vector and smaller than the current vector, one has passed the perihelion and its position is given by \code{vec\_r\_last}.
It is useful to save the position of the perihelion in a list \code{list\_perih} for a certain number of turns (here e.g.\ \code{max\_turns = 10}).
The implementation can be realized as follows\footnote{%
	Note that one stores \code{vec\_r\_last} by making a copy of \code{M.pos}: \code{vec\_r\_last = vector(M.pos)}.
	This is essential because otherwise, if one changes \code{M.pos}, one would automatically change \code{vec\_r\_last} as well.
}
\begin{lstlisting}
# Set up vectors
vec_rM_last = vec_rM0
turns      = 0
max_turns  = 10
list_perih = list()
# Find perihelion for each turn and print it out
while turns < max_turns:
    vec_rM_before_last = vec_rM_last
    # Store position of Mercury in a new vector (since we will change M.pos)
    vec_rM_last = vector(M.pos)
    #<...update Mercury position...>
    # Check if at perihelion
    if (vec_rM_last.mag < M.pos.mag) and (vec_rM_last.mag < vec_rM_before_last.mag):
        list_perih.append(vec_rM_last)
        turns = turns + 1
\end{lstlisting}
The angle between two vectors can be computed using
 \begin{equation}
 	\sphericalangle(\vec{v}_{1},\vec{v}_2) = \mathrm{arccos} \left( \frac{\vec{v}_{1} \cdot \vec{v}_2}{|\vec{v}_{1}|\:|\vec{v}_2|} \right)
	\, .
 \end{equation}
This is readily implemented in \texttt{VPython} via
\begin{lstlisting}
# Import functions to compute angle
from vpython import acos, pi, dot
# Define function for angle extraction
def angle_between(v1, v2):
    return acos( dot(v1, v2) / (v1.mag * v2.mag) ) * 180. / pi
\end{lstlisting}
where the factor $180/\pi$ in the last line converts the unit of the angle from radians into degrees.
To account for the statistical errors (e.g.\ numerical rounding errors) one can average over a few turns.
Depending on the programming proficiency of the students and time constraints, this can be done either by hand or, e.g., by implementing the following code
\begin{lstlisting}
sum_angle=0.
for n in range(1, max_turns):
    # Calculate angle
    sum_angle = sum_angle + angle_between(list_perih[n-1], list_perih[n])
# Display the average
print(sum_angle/(max_turns-1))
\end{lstlisting}
Note that the perihelion motion is computed from the locations stored in \code{list\_perih} and not based on the initial position.
This is important because depending on the initial conditions and numerical uncertainties, the simulation does not necessarily start in the exact perihelion.

As explained above [and further discussed in Section~\ref{sec:analysis}] the natural value for $\alpha$ (and $\beta$)  would be of the order of $1$.
However, if the students use this value for $\alpha$, they will find that the change in the trajectories is close to invisible and the numerical uncertainty is much larger than the result.
It is therefore more advisable to use the fact that there is a linear dependence between $\alpha$ and $\delta \Theta$ to estimate the size of the perihelion motion.
On the other hand as soon as the values of $\alpha$ or $\beta$ get too large the effective ansatz to use only the leading terms 
in the $(r_s/r)$ and $(r_L/r)^2$ expansion is no longer justified and the method of extraction might get unreliable.
At this point we therefore recommend to choose the parameters  $\alpha$ and $\beta$ at most of the order of $10^5$ --- this
value will be better justified in Section~\ref{sec:analysis}.

\begin{figure}[htb]
	\centering
	\begin{subfigure}[c]{0.49\textwidth}
		\includegraphics[width=\textwidth]{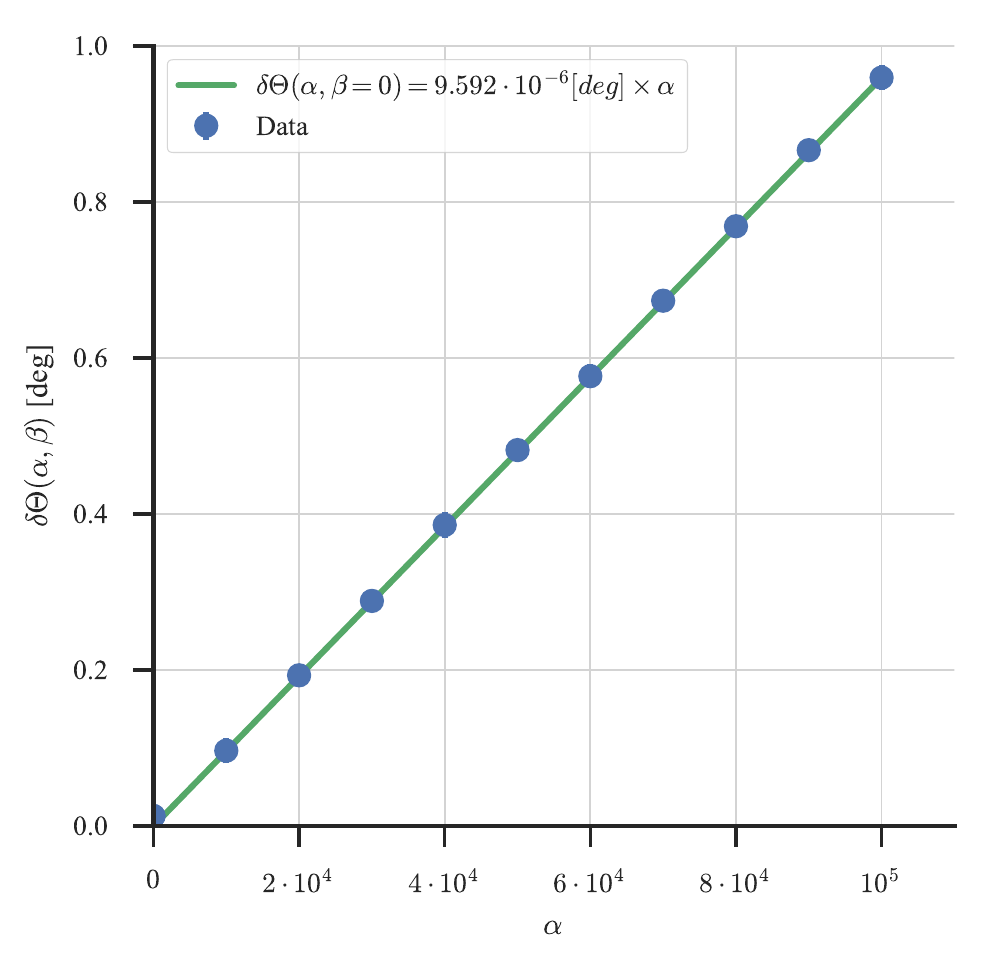}
		\caption{\label{fig:angle-alpha}}
	\end{subfigure}
	\begin{subfigure}[c]{0.49\textwidth}
		\includegraphics[width=\textwidth]{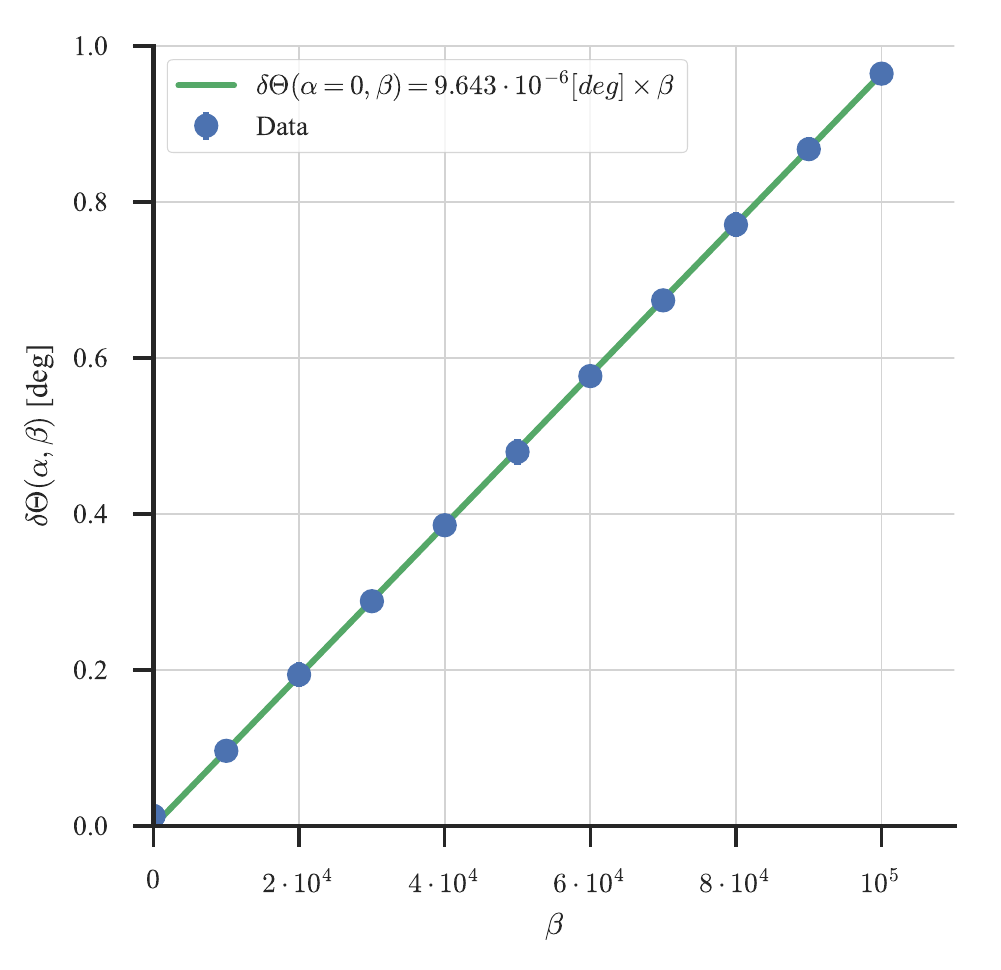}
		\caption{\label{fig:beta-alpha}}
	\end{subfigure}
	\caption{\label{fig:AlphaAngle} Linear relation between $\alpha$ (a), $\beta$ (b) and the perihelion motion $\delta \Theta$ for $\Delta t=2v_M(0)/a_M(0)/200$ and the other parameter set to zero.}
\end{figure}

The students should convince themselves that in this parameter range there is a linear relation between the angles $\delta \Theta$ and the parameters $\alpha$ or $\beta$:
\begin{equation}\label{eq:theta-linear}
	\delta\Theta (\alpha, \beta) = m_\alpha \cdot \alpha + m_\beta \cdot \beta
	\, .
\end{equation}
This can be done either by hand, \python{} (e.g., with \texttt{matplotlib}~\cite{Matplotlib}), or using another program like, e.g., Excel.
For instance Figure~\ref{fig:AlphaAngle} demonstrates such plots.

As an example we  extract the perihelion motion using the parameters provided by General Relativity, namely $\alpha = 0$ and $\beta = 3$ evaluated with $\Delta t = \Delta t_0 / 200$.
Considering one result from the numeric simulation, e.g., $\delta\Theta(\alpha = 0, \beta=10^5) = 0.964^{\degree}$, the sought-after angle can be calculated using Eq.~(\ref{eq:theta-linear})
\begin{equation}
	\delta\Theta (\alpha=0, \beta=3) = \frac{0.964^{\degree}}{10^5} \cdot 3 = 0.104''
	\, .
\end{equation}
Depending on the knowledge of the students, multiple points and also estimated numerical uncertainties could be used to extract this value by a linear regression.
Here, one has to find a proper range for the interpolation.
Small values for $\alpha$ and $\beta$ usually come with relatively larger numerical errors while for larger values one cannot guarantee that the additional terms are sufficiently small.
For this reason, we suggest to interpolate between zero and $\alpha_{\max}$, $\beta_{\max} \sim 10^5$.
This will be further explained in the next section.
Note that, when comparing the results of the simulation with the experimental result, one Earth year corresponds to $T_E/T_M \approx 365~\mathrm{d}/88~\mathrm{d}\approx 4.15$ Mercury years.
Therefore, the perihelion motion per 100 earth years is $0.104''\cdot 415\approx 43.2''$.
This is consistent with the observation $\delta\Theta_{\mathrm{M}} = (42.56\pm 0.94)''$ [see Eq.~(\ref{delT})].

This agreement is the second important achievement after which finishing the course appears natural: The students have understood quantitatively the
perihelion motion of Mercury. The next section that finalizes the numerical investigation is even more technical and should
be worked on by the most advanced students only.

\section{Tests of stability and error analysis}\label{sec:stability}
Measurements as well as numerical simulations in physics should always be accompanied by estimates of the corresponding uncertainties.
It is important for the students to recognize this fact and to understand what the sources of uncertainties are.
The students should therefore explore sources and sizes of inaccuracies arising in this simulation at least qualitatively.

There are many potential sources of uncertainties in a simulation and discussing all of them is be beyond the scope of this work.
Instead, we focus on the most accessible source: Numerical errors due to finite time steps $\Delta t$.
Additional sources of errors include the omission of terms in Eqs.~(\ref{eq:vdef}) and~(\ref{eq:adef}) and the infinite mass approximation of the Sun (i.e.\ keeping the Sun's position fixed).

\begin{figure}[htb]
	\centering
	\includegraphics[width=.99\textwidth]{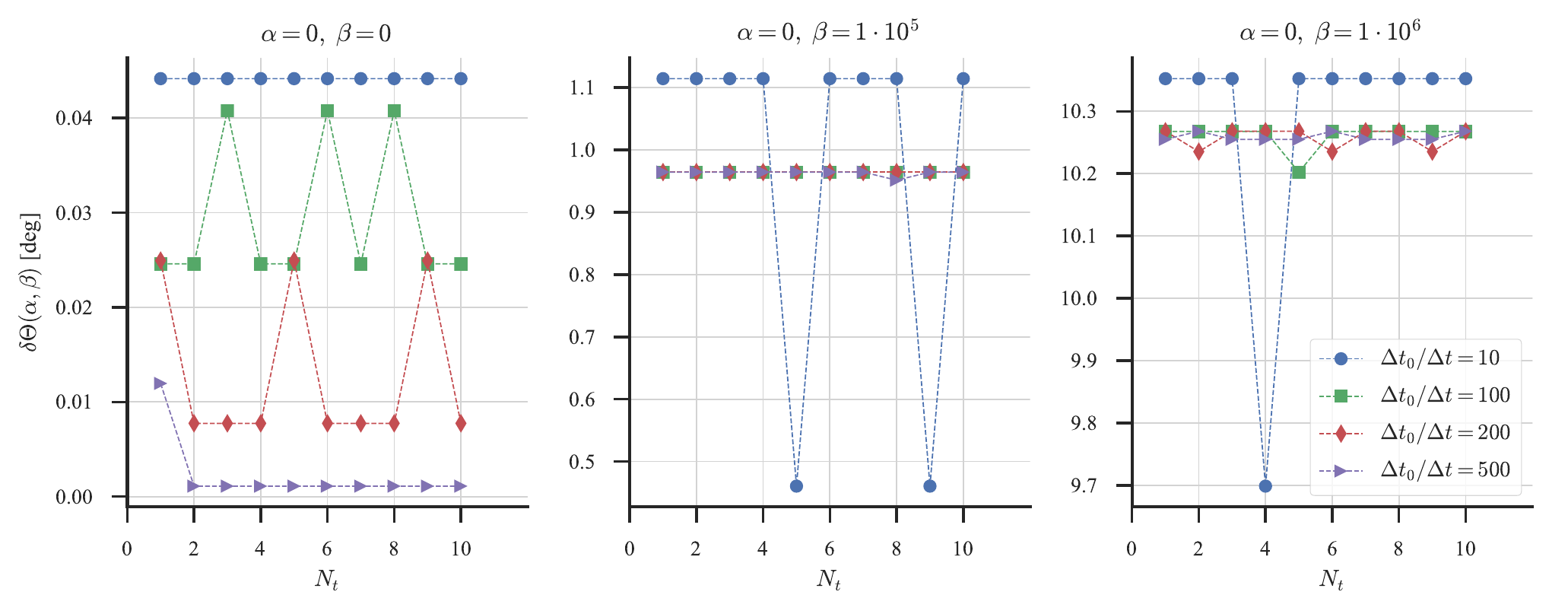}
	\caption{\label{fcc3}The motion of the perihelion $\delta\Theta$ in degrees depending on the number of turns $N_t$ for different time steps $\Delta t$ (color) and different values for $\beta$ (columns).
	Here, $\Delta t_0 \equiv 2 v_M(0)/a_M(0)$.
	Offsets of $\delta \Theta(\alpha=0, \beta=0)$ from zero can be used as an estimate of the magnitude of the error. Oscillations in $\delta \Theta$ indicate too coarse time steps. 
	Note the change in scale for the different plots.
}
\end{figure}

Consider Figure~\ref{fcc3}, where the angle of the perihelion motion $\delta \Theta$ is shown for the first $10$ turns ($x$-axis) for different choices of $\Delta t$ (colors) and $\beta$ (columns). In all cases $\alpha=0$.
As expected, $\delta\Theta$ is approximately constant for sufficiently small time-steps $\Delta t$ and its value depends solely on $\beta$.
However, contrary to the correct result $\delta\Theta$ deviates from zero for $\alpha = 0 = \beta$ for all $\Delta t$.
Using the data points in Figure~\ref{fig:AlphaAngle} and extrapolating to $\alpha = 0 = \beta$ without enforcing $\delta\Theta(0,0) = 0$ leads to a similar offset.
This offset is a numerical error and can be taken as an estimate for the error of the simulation.

Furthermore, one can observe that for large time-steps the values of the perihelion motion oscillate between discrete values at different turns.
This can be explained as follows: The sample code we present to find the perihelion vectors only finds the position closest to the Sun amongst
the discrete set of vectors evaluated at the discretized time steps and labels it as perihelion vector. However,
this is only an approximation: Sometimes the program finds a point before and sometimes after the perihelion causing the oscillations observed.
The precision of this approximation improves as the time-step shrinks.
Hence, strongly visible oscillations in $\delta\Theta$ for a different number of turns indicate that the time-step used is too large.

This problem can be pointed out to the students by asking them to vary the size of the time-steps and observe the effect on the trajectories.
This should also be done for a wide range of values to show both good and problematic regimes.
It is advantageous to set both $\alpha$ and $\beta$ to zero for this study, since in this case the trajectories should be closed and deviations from the correct case can be spotted easily.
Figure~\ref{fig:MercuryOrbit-a0-small-dt-large} shows the Mercury trajectory for $\Delta t=2 v_M(0) /a_M(0) \times 2$.
One can clearly see the failure of the simulation to reproduce the physical trajectory shown as the black, solid line.
From examples like this it should be apparent that the time steps influence the accuracy of the simulation.
The simulation reproduces the actual trajectory only in the limit $\Delta t \rightarrow 0$. Thus, in any numerical simulation
one always has to identify a proper compromise between numerical accuracy and time spent for the simulation
(clearly for $\Delta t\to 0$ the computing time goes to infinity).

As a first estimate, one can approximate the numerical uncertainty of the perihelion motion, $\Delta \delta \Theta (\alpha, \beta)$ at non-zero $\alpha$ and $\beta$, by its offset
 at $\alpha = 0 = \beta$,  the amplitude of its oscillation at $\alpha = 0 = \beta$ as well as the amplitude of its oscillation at the non-zero values of $\alpha$ and $\beta$
 used for the actual calculation.
\begin{eqnarray}\label{eq:num-uncertainty}
	\Delta \delta \Theta (\alpha, \beta) &=& \sqrt{
		\delta \Theta_{mean}^2 (0, 0) + 
		\delta \Theta_{std}^2 (0, 0) + 
		\delta \Theta_{std}^2 (\alpha, \beta)
	} \, ,
	\\
	\delta \Theta_{mean} (\alpha, \beta)
	&=& 
	\frac{1}{N_{t}}
	\sum\limits_{n=1}^{N_{t}} \delta \Theta_{n} (\alpha, \beta) \, ,
	\\
	\delta \Theta_{std}^2 (\alpha, \beta)
	&=& 
	\frac{1}{N_{t}-1}
	\sum\limits_{n=1}^{N_{t}} \left[
		\delta \Theta_{mean} (\alpha, \beta) - \delta \Theta_{n} (\alpha, \beta)
	\right]^2 \, ,
\end{eqnarray}
where the standard deviations and mean values are obtained from the data sample of $N_t$ turns (orbits of mercury).
Thus, the numerical value extracted is given by
\begin{equation}\label{eq:num-vals}
	\delta \Theta (\alpha, \beta) = \delta \Theta_{mean} (\alpha, \beta)  \pm \Delta \delta \Theta (\alpha, \beta)
	\, .
\end{equation}
Note that this has to be repeated for each different time step $\Delta t$.

\begin{figure}[htb]
	\centering
	\includegraphics[width=.99\textwidth]{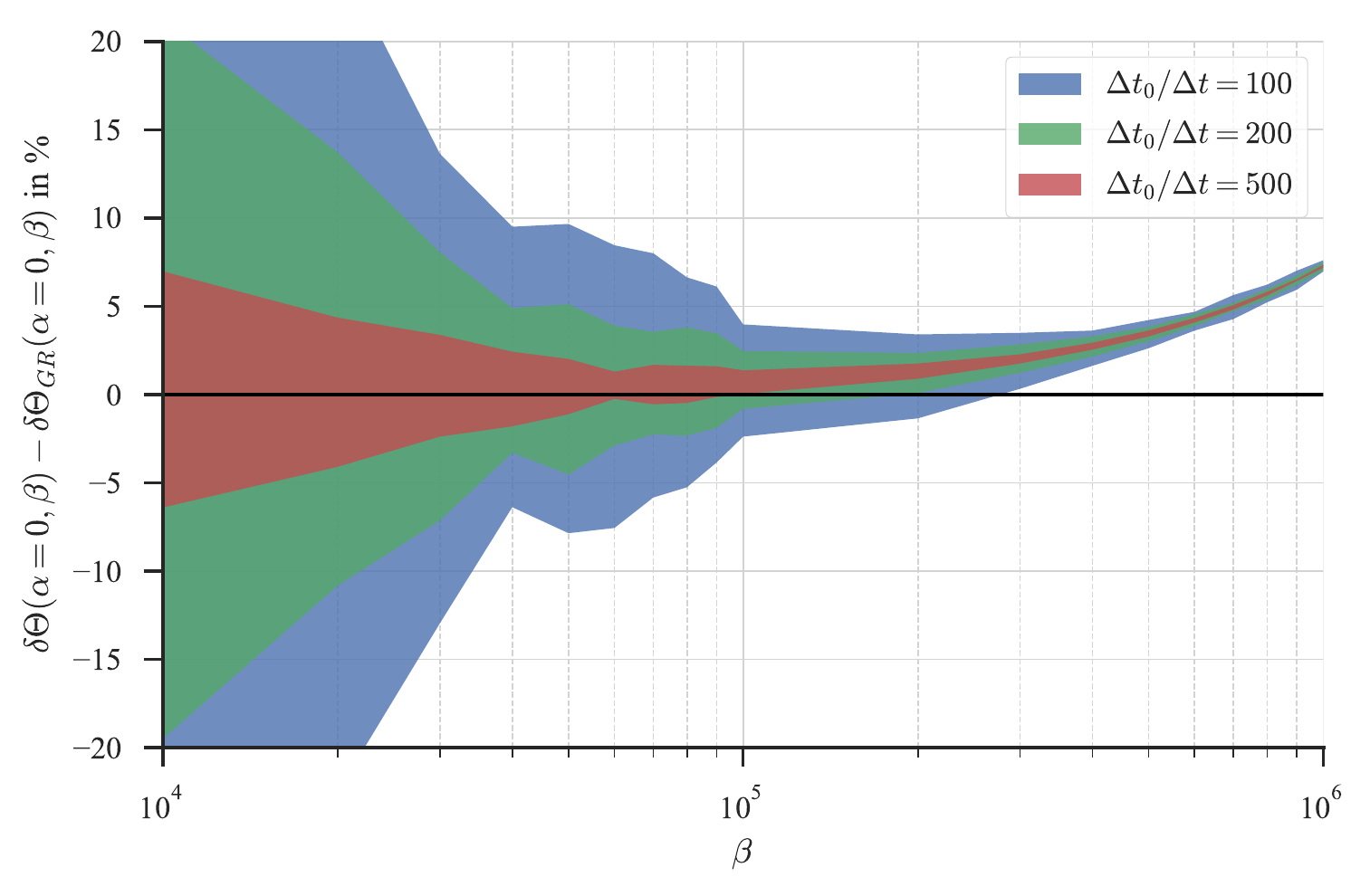}
	\caption{\label{fcc4}Difference of numerically extracted perihelion motion relative to the value computed from General Relativity [Eq.~(\ref{eq:gr-prediction})] for different time steps.  The error bands are computed according to Eq.~(\ref{eq:num-uncertainty}).
}
\end{figure}

The absolute value of the numerical error mostly depends on the time step $\Delta t$ and is roughly independent of the values of $\alpha$ and $\beta$.
Therefore, it is more desirable to pick large values for $\alpha$ and $\beta$, because this increases the absolute size of the perihelion
motion and thus decreases the relative numerical error.
However, there is a competing effect which places upper bounds on the values of $\alpha$ and $\beta$.
For instance, the prediction for the perihelion motion coming from General Relativity allowing for varying values of $\beta$ is given by
\begin{equation}\label{eq:gr-prediction}
	\delta \Theta_{GR} (\beta) = 
	2 \pi \left[
	\frac{\beta}{4} \,\frac{r_S^2}{r_L^2}
	+\mathcal{O}\left(\beta^2\,\frac{r_S^4}{r_L^4}\right)
	\right]
	\, .
\end{equation}
Thus, for large values of $\beta$, quadratic corrections in $\beta$ become relevant and it is not possible to extract the 
perihelion growth of Mercury by performing a linear interpolation.
We display these competing effects in Figure \ref{fcc4}.
The relative difference between numerically extracted values and the General Relativity prediction for $\delta \Theta$ are plotted against $\beta$ for zero $\alpha$.
The value of $\beta$ we recommend for the extraction is of the order of $10^5$ as will be further motivated in the next section.
As can be seen in the figure: For smaller values of $\beta$, the relative numerical uncertainty grows, while for larger 
 values, $\beta \gtrsim 10^5$, the numerical values deviate from the assumed linear dependence on $\beta$.
Identifying a parameter space for reliable and precise computations is a general challenge for numerical simulations.

\section{Dimensional analysis}\label{sec:analysis}

Dimensional analysis is not only a tool that allows one to cross check, if the results of simulations are of the right order of magnitude, it is also very helpful to identify unusual dynamics in some systems.
Especially the latter aspect should become clear from the discussion in this section. The modifications to the Newtonian equation of gravity introduced to account for the perihelion motion of Mercury
are discussed also in Ref.~\cite{Wells:2011st}. To get a deeper understanding of the concepts introduced in this section reading this article is highly recommended.

The idea of dimensional analysis is that in a system that can be controlled by expanding the relevant quantities (like the force) in some small parameter(s), the coefficients in the expansion should turn out to be of order one (that means anything between about 0.1 and 10 is fine --- but 0.01 or 100 is irritating); parameters in line with this are called "natural".
Applied to the problem at hand given by Eq.~(\ref{eq:newton}) this statement implies that from naturalness one would predict the parameters $\alpha$ and $\beta$ to be
of order one.
One expects that unnatural parameters point at dynamics not accounted for explicitly.
Employing for the problem at hand the average distance Mercury-Sun $\bar r_{MS}=6\times 10^7\,\mathrm{km}$, the concept on 
naturalness allows us to estimate the expected angular shift per orbit for, e.g., $\alpha=1$ and $\beta=0$
\begin{equation}
\delta \Theta \simeq 2\pi\left(\frac{r_S}{\bar r_{MS}}\right) = (\pi \times 10^{-7}) \ \mbox{rad} = (2\cdot 10^{-5})^o = (7\cdot 10^{-2}) ^{''} \ ,
\end{equation}
or for $\alpha=0$ and $\beta=1$
\begin{equation}
\delta \Theta \simeq 2\pi\left(\frac{r_L^2}{\bar r^2_{MS}}\right) = (\pi \times 5 \cdot 10^{-8}) \ \mbox{rad} = (1 \cdot 10^{-5})^o = (4\cdot 10^{-2}) ^{''} \ ,
\end{equation}
which leads to a shift of about $30^{''}$ and $15^{''}$, respectively, in 100 earth years. These are to be compared to the empirical value of $43^{''}$.
Thus the amount of perihelion motion of Mercury is indeed in line with expectations, \textit{if} --- and this is an important "if" --- the 
Newtonian dynamics is simply the leading term of some more general underlying theory. 
 In particular, no new scales enter in the correction terms in addition to $r_S$ and $r_L^2$. This is by itself already an
 interesting observation.
 In case of General Relativity $\beta=3$ is a natural value 
while $\alpha$ vanishes. This pattern is therefore a non-trivial prediction of General Relativity and one should expect
that alternative theories of gravitation generate non-vanishing values of $\alpha$. 

Since the dimensional analysis allows one to estimate with little effort a certain effect to be studied in numerical simulation
one may also use it as a check of the numerical results: If the simulation had produced a result orders of magnitude different
from the expectations of the dimensional analysis, there must be a dynamical reason in the physical system that deserves to be identified --- or the code underlying the simulation has an error. 

It is even possible to push the idea of naturalness further to estimate the intrinsic uncertainty of a given study. 
For the problem at hand we identified the expansion parameters $(r_s/r)$ and $(r_L/r)^2$ both being of order $10^{-7}$.
Therefore, as long as natural parameters are employed in the simulation one expects corrections to be suppressed by
seven orders of magnitude since those need to scale as $(r_s/r)^2$, $(r_L/r)^4$, or $(r_s r_L^2/r^3)$.
In the simulations discussed above we observed that for $\beta=10^6$ the deviation of the result from the expectation
of the underlying theory is 8\% ($cf$. Fig.~\ref{fcc4}). Even this is in line the estimate just discussed, since 
for such large values of $\beta$ the effective expansion parameter is $\beta(r_L/r)^2\sim 10$\%. This is the 
justification for limiting $\beta$ for a reliable extraction of the perihelion motion of Mercury to values below $10^5$.

Please note, however, dimensional analysis only provides an order of magnitude estimate of a given effect. 
While it can be used to cross check some explicit detailed evaluation it can by no means replace it, if one
aims at precise results.

We do not want to leave unmentioned that in modern physics the concept of naturalness plays a very important role. 
It is regarded as a serious problem, when parameters deviate significantly from their natural values. 
Nowadays there are  several of those hierarchy problems in modern physics: e.g., the so called QCD $\Theta$ term, 
expected to be of natural size, is at present known to be at most $10^{-10}$.
This smallness, called the strong CP problem, is so irritating that physicists like S. Weinberg even proposed that there must exist an additional particle, the axion.
Its interactions would even push $\Theta$ to zero.
At present various intense experimental searches for this axion are going on at various labs.

\section{Possible extensions}\label{sec:extensions}

\begin{itemize}

\item \textbf{Explore problem autonomously}

In Section~\ref{sec:Numerical Implementation} we suggested to present the material by using a template as well as a step by step instruction to guide the students through the problem solution \cite{scripts}.
These instructions can be cut down or left out depending on available time and numerical/computational versatility of the students.
This could be achieved by the following changes or additions to the concept presented in Section~\ref{sec:Numerical Implementation}.

\begin{itemize}
\item \textbf{Build code from scratch}

Instead of providing the template to the students, they could build the program from scratch.
Of course, this requires some basic knowledge in \vpython{}, which they could acquire for example by working through introductory materials (see~\cite{VPython}).
Also, they could independently research the parameters relevant for the system.
\item \textbf{Why can we work in a plane?}

In the code the third coordinate of all vectors is set to zero and never used.
On the first glance, this might seem like a simplification.
This choice is however possible without loss of generality because we are dealing with a central potential.
The students could work out the reason behind this choice on their own.
\item \textbf{What is the impact of the different parameters?}

Especially if the parameters are not specified beforehand, the students might have to experiment a bit, before getting the correct trajectories.
But even if they are given, it might be beneficial to encourage the students to play with a few parameters, like the masses or the starting velocities, and observe their impact on the trajectories.
This way the students get a better feeling for the physics involved.
\end{itemize}

\item \textbf{Optimizing performance}

Simulations always involve a balance between time needed for the calculations and the demanded accuracy for the results.
Even though this is a rather simple example, it contains some opportunities to make these concepts accessible to the students.

\begin{itemize}
\item \textbf{Measure calculation time}

In practice there is a limit to decreasing the time steps, because the time needed for the calculation grows simultaneously.
The following snippet shows how the run time of function \texttt{main} can be measured:
\begin{lstlisting}
import time
start_time = time.time()
main()
print("--- {} seconds ---".format(time.time() - start_time))
\end{lstlisting}
By varying $\Delta t$ the students can validate that there is indeed approximately an anti-proportional dependence.
(Note: This only works if the time in the loop is increased by $\Delta t$, so $t=t+\Delta t$.)
\item \textbf{Verlet algorithm}

Obviously, an improvement in accuracy can be achieved by using a better algorithm without changing $\Delta t$.
The simplest way to demonstrate this might be given by the implementation of the Verlet algorithm (see e.g.~\cite{Hairer03geometricnumerical} and references therein) instead of using the simple Euler method employed here for solving the differential equation.
\end{itemize}

\item \textbf{Extended problems}

\begin{itemize}
\item \textbf{Non-stationary Sun}

It might be interesting to abandon the simplification of a stationary Sun, as it nicely illustrates Newton's third law.
Here it might also be advisable to reduce the mass of the Sun to have a more visible result.

\item \textbf{Three-body problem}

Ambitious students could even include further planets and see how the different planets interact.
This is especially interesting, when discussing the perihelion motion of Mercury, as it is mainly due to the influence of the other planets.
Only a smaller part is due to General Relativity.

\end{itemize}

\end{itemize}

\section{Summary}

In this work a course is presented that should enable advanced high school students to understand quantitatively the perihelion motion of Mercury
by using a numerical simulation. At the same time the active participation in the course teaches the central role of 
differential equations in theoretical physics, the basic concepts of how to use numerical simulations to find their
solutions as well as the need to estimate the uncertainties of a given study.
In addition the concepts of dimensional analysis and naturalness were introduced which not only allow for an
estimate of a given effect a priori (to cross check the numerical results) but also to estimate the intrinsic uncertainty
of the result.

The course is structured such that students at different levels can stop after different achievements. The basic
course contains the set-up of the numerical simulation and its visualization: The students will have observed
the impact of different forces on the trajectories and some basic features of numerical studies.
The more advanced students may proceed to extract the perihelion motion quantitatively from the 
parameters provided by General Relativity. And finally the most advanced students may even follow the 
discussion of the uncertainties of the simulation. 

We are convinced that the course presented in this paper is very well suited to teach high school students not
only the power of numerical simulations but also the beauty of theoretical physics.

\section*{Acknowledgments}

This work is supported in part by NSFC and DFG through funds provided to the
Sino--German CRC110 ``Symmetries and the Emergence of Structure in QCD''.


\appendix
\section*{References}
\bibliographystyle{unsrt}
\bibliography{References}

\clearpage
\section{The code covered by the basic course (visualization of the trajectories)}\label{sec:code}
Below we provide the complete code as well as a flowchart for part 1 as developed above.
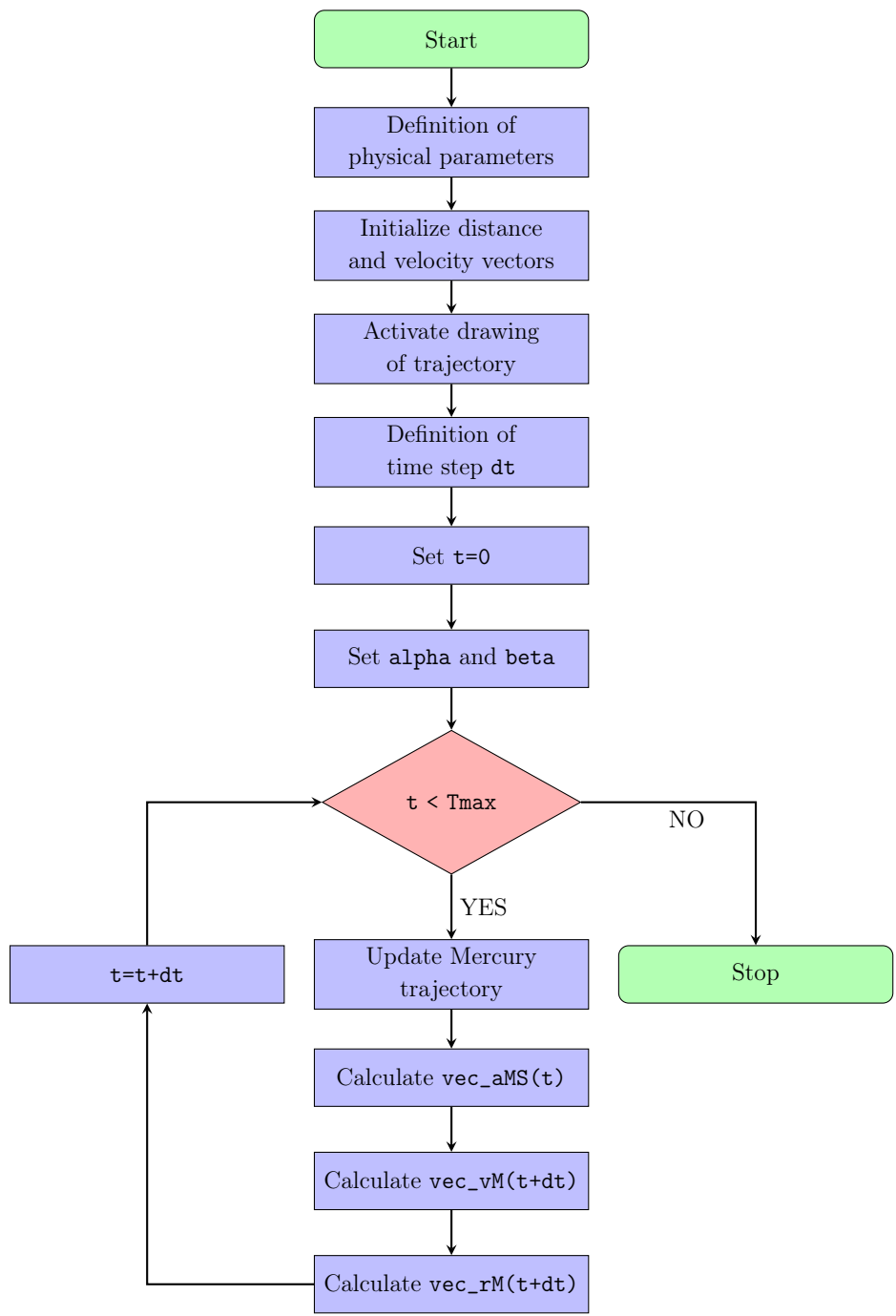
\begin{figure}[htb]
\centering
\begin{tikzpicture}[node distance=1.8cm, every node/.style={scale=0.8}]
  \node (start)  [startstop] {Start};
  \node (proc1)  [process, below of=start] {Definition of\\ physical parameters};
  \draw [arrow]  (start) -- (proc1);
  \node (proc2)  [process, below of=proc1] {Initialize distance and velocity vectors};
  \draw [arrow]  (proc1) -- (proc2);
  \node (proc4)  [process, below of=proc2] {Activate drawing of trajectory};
  \draw [arrow]  (proc2) -- (proc4);
  \node (proc5)  [process, below of=proc4] {Definition of time step \texttt{dt}};
  \draw [arrow]  (proc4) -- (proc5);
  \node (proc6)  [process, below of=proc5] {Set \texttt{t=0}};
  \draw [arrow]  (proc5) -- (proc6);
  \node (proc6b) [process, below of=proc6] {Set \texttt{alpha} and \texttt{beta}};
  \draw [arrow]  (proc6) -- (proc6b);
  \node (dec1)   [decision, below of=proc6b, yshift=-0.7cm] {\texttt{t < Tmax}};
  \draw [arrow]  (proc6b) -- (dec1);
  \node (proc7)  [process, below of=dec1, yshift=-1.2cm] {Update Mercury trajectory};
  \draw [arrow]  (dec1) -- node[anchor=west] {YES} (proc7) ;
  \node (proc8)  [process, below of=proc7] {Calculate \texttt{vec\_aMS(t)}};
  \draw [arrow]  (proc7) -- (proc8);
  \node (proc9)  [process, below of=proc8] {Calculate \texttt{vec\_vM(t+dt)}};
  \draw [arrow]  (proc8) -- (proc9);
  \node (proc10) [process, below of=proc9] {Calculate \texttt{vec\_rM(t+dt)}};
  \draw [arrow]  (proc9) -- (proc10);
  \node (stop)   [startstop, right of=proc7, xshift=3.5cm] {Stop};
  \node (proc11) [process, left of=proc7, xshift=-3.5cm] {\texttt{t=t+dt}};
  \draw [arrow]  (dec1) -| node[anchor=north, xshift=-1.2cm] {NO} (stop);
  \draw [arrow]  (proc10) -| (proc11);
  \draw [arrow]  (proc11) |- (dec1);
\end{tikzpicture}
\caption{\label{fig:flow}Flowchart demonstrating the logical ordering of the example code.}
\end{figure}

\begin{figure}[htb]
\begin{lstlisting}[language=Python,basicstyle=\linespread{1}\footnotesize\ttfamily]
# Import all objects from the vpython module
from vpython import *

# Definition of physical parameters
rM0 = 4.60    # Initial radius of Mercury orbit, in units of R0
vM0 = 5.10e-1 # Initial orbital speed of Mercury, in units of R0/T0
c_a = 9.90e-1 # Base acceleration of Mercury, in units of R0**3/T0**2
TM  = 8.80e+1 # Orbit period of Mercury
rS  = 2.95e-7 # Schwarzschild radius of Sun,in units of R0
rL2 = 8.19e-7 # Specific angular momentum, in units of R0**2

# Define the initial coordinates; M = mercury, S = Sun
M = sphere(pos=vector(0, rM0, 0), radius=0.5,  color=color.red   )
S = sphere(pos=vector(0, 0, 0),   radius=1.5,  color=color.yellow)
# And the initial velocities
M.velocity = vector(vM0, 0, 0)
S.velocity = vector(0, 0, 0)

# Add a visible trajectory to mercury
M.trajectory = curve(color=color.white)

# Definition of the time step
dt = 2 * vM0 / c_a / 20

# Define the coordinate and velocity update
def evolve_mercury(vec_rM_old, vec_vM_old, alpha, beta):
    # Compute the strength of the acceleration
    temp = 1 + alpha * rS / vec_rM_old.mag + beta * rL2 / vec_rM_old.mag**2
    aMS  = c_a * temp / vec_rM_old.mag**2
    # Multiply by the direction
    vec_aMS = - aMS * ( vec_rM_old / vec_rM_old.mag )
    # Update velocity vector
    vec_vM_new = vec_vM_old + vec_aMS * dt
    # Update position vector
    vec_rM_new = vec_rM_old + vec_vM_new * dt
    return vec_rM_new, vec_vM_new

t     = 0.0
alpha = 0.0
beta  = 0.0
# Execute the loop as long as t < 2*TM
while t < 2*TM:
    # Set the frame rate (you can choose a higher rate to accelerate the program)
    rate(100)
    # Update the drawn trajectory with the current position
    M.trajectory.append(pos=M.pos)
    # Update velocity and position
    M.pos, M.velocity = evolve_mercury(M.pos , M.velocity , alpha, beta)
    # Advance time by one step
    t = t + dt
\end{lstlisting}
\end{figure}
Further examples and template files can be found online \cite{scripts}.

\end{document}